# THE PECULIARITIES OF NONSTATIONARY FORMATION OF INHOMOGENEOUS STRUCTURES OF CHARGED PARTICLES IN THE ELECTRODIFFUSION PROCESSES

Nefyodov P., Reztsov V., Riabinina O.

Ukraine, Kiev

In this paper the distribution of charged particles is constructed under the approximation of ambipolar diffusion. The results of mathematical modelling in two-dimensional case taking into account the velocities of the system are presented.

Charge transfer processes play an important part in the formation of inhomogeneous charged particle concentration profiles in various electrochemical systems [1]. Approaches to the investigation of the formation of charged particles profiles remain the topical [2,3]. There are papers in which the ion distribution is modeled in the approximation of ambipolar diffusion [4]. On the other hand, related problems are important for the simulation of processes in semiconductor devices. [5]

Consider the charged particles system, characterized by the macroscopic velocity $v^*$.

The complete system of electodiffusion equations in two-dimensional case, neglecting outer magnetic fields, could be written as follows:

$$\frac{\partial N_e}{\partial t} + v_x^* \frac{\partial N_e}{\partial x} + v_y^* \frac{\partial N_e}{\partial y} - D_e \left( \frac{\partial^2 N_e}{\partial x^2} + \frac{\partial^2 N_e}{\partial y^2} \right) - \mu_e \left( \frac{\partial (N_e E_x)}{\partial x} + \frac{\partial (N_e E_y)}{\partial y} \right) = $$
$$= \alpha_i N N_e - \alpha_r N_e^2 N_i; \quad (1)$$

$$\frac{\partial N_i}{\partial t} + v_x^* \frac{\partial N_i}{\partial x} + v_y^* \frac{\partial N_i}{\partial y} - D_i \left( \frac{\partial^2 N_i}{\partial x^2} + \frac{\partial^2 N_i}{\partial y^2} \right) + \mu_i \left( \frac{\partial (N_i E_x)}{\partial x} + \frac{\partial (N_i E_y)}{\partial y} \right) = $$
$$= \alpha_i N N_e - \alpha_r N_e^2 N_i; \quad (2)$$

$$\frac{dE_x}{dx} + \frac{dE_y}{dy} = 4\pi e (N_e - N_i), \quad (3)$$

where $N_e$, $N_i$, $N$ are the three-dimensional concentrations of positive, negative particles and neutrals; $E_x$, $E_y$, $v_x^*$, $v_y^*$ are the components of the local electrical field strength and the macroscopic movement velocity, respectively; $e$ is the electron charge; $D_e$, $D_i$ are the diffusion coefficients of positive and negative particles, which are connected with the movabilities $\mu_e$, $\mu_i$ by Einstein's relation; $\alpha_i$, $\alpha_r$ are the ionization and the recombination coefficients of the process

$$\overline{e} + \overline{e} + i \leftrightarrow a + \overline{e}. \quad (4)$$

Let us assume that the effective temperatures of positive and negative particles are equal, neutrals concentration $N$ is space- and time-constant, and introduce the quasineutrality parameter $k' = \dfrac{N_e - N_i}{N_e}$. Then, using (1) and (2), we obtain the combined equations for $N_e$, the electric field $E$ and velocity $v^*$:

$$\frac{\partial N_e}{\partial t} + v_x^* \frac{\partial N_e}{\partial x} + v_y^* \frac{\partial N_e}{\partial y} - D_e \left( \frac{\partial^2 N_e}{\partial x^2} + \frac{\partial^2 N_e}{\partial y^2} \right) - \mu_e \left( \frac{\partial (N_e E_x)}{\partial x} + \frac{\partial (N_e E_y)}{\partial y} \right) = \\ = \alpha_i N N_e - \alpha_r (1-k') N_e^3; \quad (5)$$

$$\frac{\partial ((1-k')N_e)}{\partial t} + v_x^* \frac{\partial ((1-k')N_e)}{\partial x} + v_y^* \frac{\partial ((1-k')N_e)}{\partial y} - D_i \left( \left( \frac{\partial^2}{\partial x^2} + \frac{\partial^2}{\partial y^2} \right)((1-k')N_e) \right) + \\ + \mu_i \left( \frac{\partial ((1-k')N_e E_x)}{\partial x} + \frac{\partial ((1-k')N_e E_y)}{\partial y} \right) = \alpha_i N N_e - \alpha_r (1-k') N_e^3; \quad (6)$$

The $|k'| \ll 1$ condition should be supplemented with the inequalities

$$\left| \frac{\partial (k' N_e)}{\partial t} \right| \ll \left| \frac{\partial N_e}{\partial t} \right|; \quad \left| \frac{\partial^2 (k' N_e)}{\partial x^2} \right| \ll \left| \frac{\partial^2 N_e}{\partial x^2} \right|; \quad \left| \frac{\partial^2 (k' N_e)}{\partial y^2} \right| \ll \left| \frac{\partial^2 N_e}{\partial y^2} \right|; \\ \left| \frac{\partial (k' N_e E_x)}{\partial x} \right| \ll \left| \frac{\partial (N_e E_x)}{\partial x} \right|; \quad \left| \frac{\partial (k' N_e E_y)}{\partial y} \right| \ll \left| \frac{\partial (N_e E_y)}{\partial y} \right|; \\ \left| \frac{\partial (k' N_e)}{\partial x} \right| \ll \left| \frac{\partial N_e}{\partial x} \right|; \quad \left| \frac{\partial (k' N_e)}{\partial y} \right| \ll \left| \frac{\partial N_e}{\partial y} \right| \quad (7)$$

The limitations on the space and time derivatives of charge density may be obtained from (7), and the combined equations for $N_e$, $E$ and $v^*$ result from (6) and (7):

$$\frac{\partial N_e}{\partial t} - \frac{D_e \mu_i + D_i \mu_e}{\mu_e + \mu_i} \cdot \left( \frac{\partial^2 N_e}{\partial x^2} + \frac{\partial^2 N_e}{\partial y^2} \right) + v_x^* \frac{\partial N_e}{\partial x} + v_y^* \frac{\partial N_e}{\partial y} = \alpha_i N N_e - \alpha_r N_e^3; \quad (8)$$

$$\frac{\partial (N_e E_x)}{\partial x} + \frac{\partial (N_e E_y)}{\partial y} = \\ = \frac{D_i - D_e}{D_e \mu_i + D_i \mu_e} \frac{\partial N_e}{\partial t} - \frac{D_i - D_e}{D_e \mu_i + D_i \mu_e} \left( \alpha_i N N_e - \alpha_r N_e^3 \right) + \frac{D_i - D_e}{D_e \mu_i + D_i \mu_e} \left( v_x^* \frac{\partial N_e}{\partial x} + v_y^* \frac{\partial N_e}{\partial y} \right). \quad (9)$$

It is evident from (8), (9) that the approximation of ambipolar diffusion allows us to determine the electron concentration profile $N_e$ at any time regardless of the local field strengths $E_x$ and $E_y$.

Let us consider further the solution of the initial-boundary problem for (8) with the Dirichlet boundary conditions on $(0, \Lambda) \times (0, \Lambda)$

$$N_e(x=0, y) = N_e(0, y), \quad N_e(x=\Lambda, y) = N_e(\Lambda, y), \\ N_e(x, y=0) = N_e(x, 0), \quad N_e(x, y=\Lambda) = N_e(x, \Lambda) \quad (10)$$

for all $t$. Initial conditions, i.e. $x$ and $y$ distribution of $N_e$

$$N_e(x, y, t=0) = N_e^*(0), \quad (11)$$

are chosen for simplicity to be the same as the boundary conditions

$$N_e^*(x=0, y) = N_e(0, y), \qquad N_e^*(x=\Lambda, y) = N_e(\Lambda, y),$$
$$N_e^*(x, y=0) = N_e(x,0), \qquad N_e^*(x, y=\Lambda) = N_e(x, \Lambda).$$
(12)

Thus, the problem of the space-time distribution determination of the electron concentration in plane comes to the solution of (8) with boundary conditions (10) and initial conditions (12). To facilitate the development of the numerical algorithm and the subsequent analysis of results it is convenient to reduce (8) to a dimensionless form:

$$\frac{\partial u}{\partial \tau} - \frac{\partial^2 u}{\partial \xi_1^2} - \frac{\partial^2 u}{\partial \xi_2^2} + v_1 \frac{\partial u}{\partial \xi_1} + v_2 \frac{\partial u}{\partial \xi_2} = \varphi(u - u^3) \; ; \qquad (13)$$

$$u(\xi_1 = 0, \xi_2) = u_{01}, \quad u(\xi_1, \xi_2 = 0) = u_{10},$$
$$u(\xi_1 = \delta, \xi_2) = u_{\delta 1}, \quad u(\xi_1, \xi_2 = \delta) = u_{1\delta}, \qquad (14)$$

$$u(\tau = 0) = u^*(\xi_1, \xi_2), \qquad (15)$$

introducing the following dimensionless constants and variables:

$$u = \frac{N_e}{N_e(\infty)}; \quad \xi_1 = \frac{x}{x^*}; \quad \xi_2 = \frac{y}{y^*}; \quad \tau = \frac{t}{t^*};$$

$$u_{01} = \frac{N_e(0, y)}{N_e(\infty)}; \quad u_{\delta 1} = \frac{N_e(\Lambda, y)}{N_e(\infty)};$$

$$u_{10} = \frac{N_e(x, 0)}{N_e(\infty)}; \quad u_{1\delta} = \frac{N_e(x, \Lambda)}{N_e(\infty)}; \qquad (16)$$

$$v_1 = \frac{v_x^*}{D_i^2 / \alpha_i}; \quad v_2 = \frac{v_y^*}{D_i^2 / \alpha_i}$$

$$u^*(\xi_1, \xi_2) = \frac{N_e^*(x, y)}{N_e(\infty)}; \quad N_e(\infty) = \left(\frac{\alpha_i N}{\alpha_r}\right)^{\frac{1}{2}};$$

$$\varphi = \alpha_i N \frac{(D_i \mu_e + D_e \mu_i) \alpha_i^2}{(\mu_e + \mu_i) D_i^4}; \quad x^* = y^* = \frac{(D_i \mu_e + D_e \mu_i) \alpha_i}{(\mu_e + \mu_i) D_i^2}; \quad t^* = \frac{(D_i \mu_e + D_e \mu_i) \alpha_i^2}{(\mu_e + \mu_i) D_i^4}.$$

The results of numerical solution of (13)-(15) are displayed in *Fig.1-2*.
The figures presented below show an influence of velocity relative to the static system ($v_1=0$ and $v_2=0$) case. The "wash-out" of spatial density along the velocity direction typical for the non-zero velocity case is noticeable. This "wash-out" increases with the increase of velocity. One should note that the presence of macroscopic velocity enhances the static density inhomogeneity near the space boundaries.

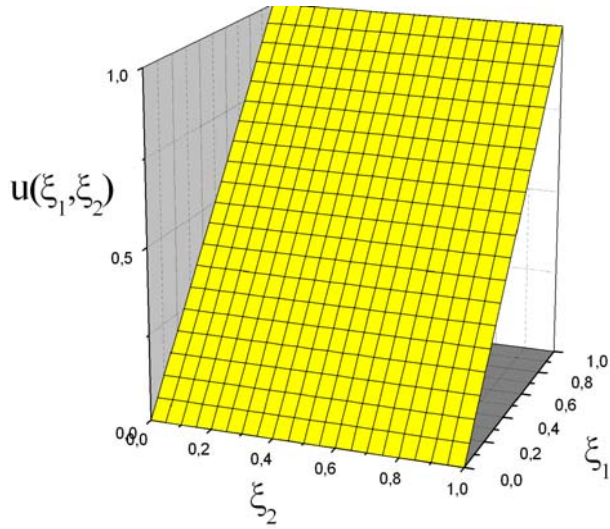

*Fig.1a* Initial linear distribution of the negative charged particles concentration.

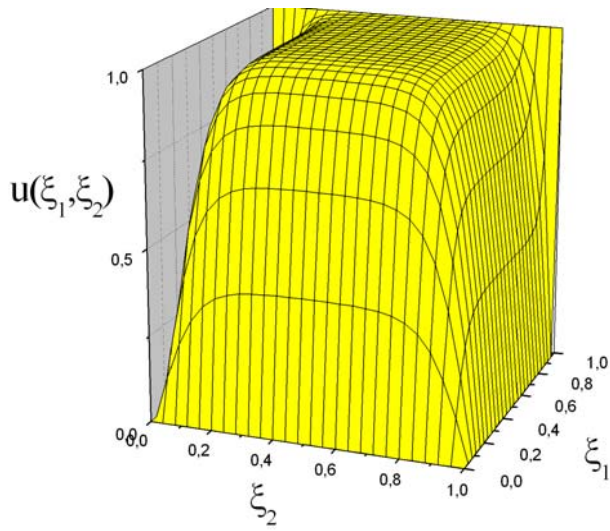

*Fig.1b* Steady-state distribution of the negative charged particles concentration ($v_1=0$ and $v_2=0$).

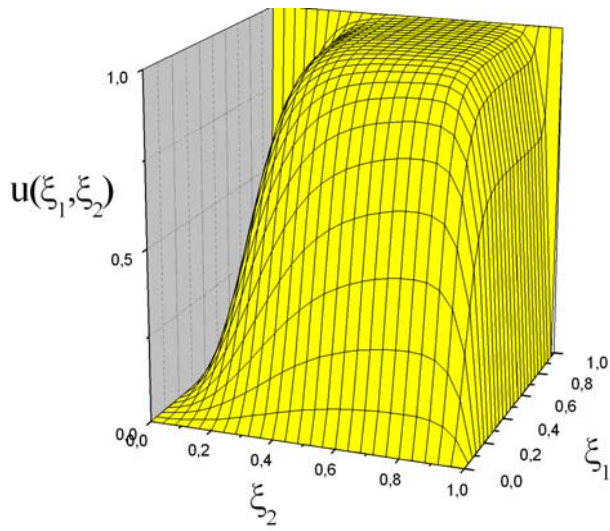

*Fig.1c* Steady-state distribution of the negative charged particles concentration ($v_1=3$ and $v_2=3$).

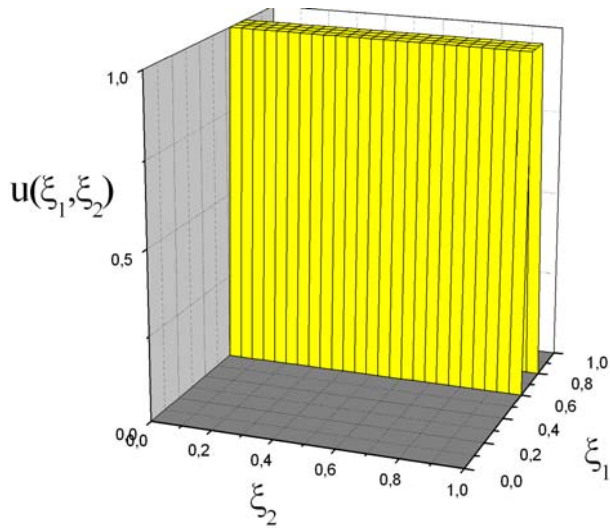

*Fig.2a* Initial plain distribution of the negative charged particles concentration.

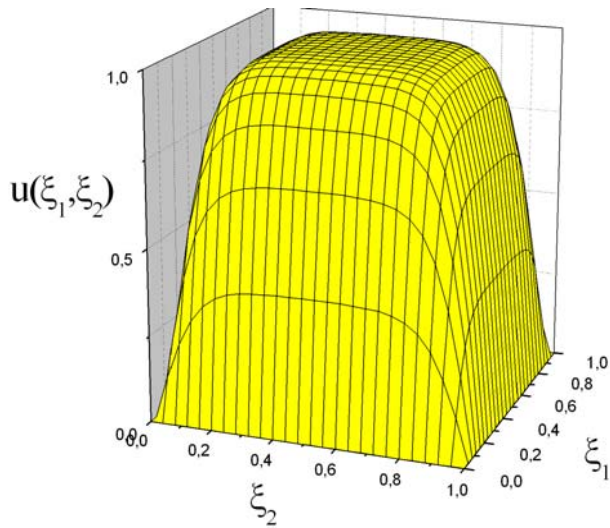

*Fig.2b* Steady-state distribution of the negative charged particles concentration ($v_1=0$ and $v_2=0$).

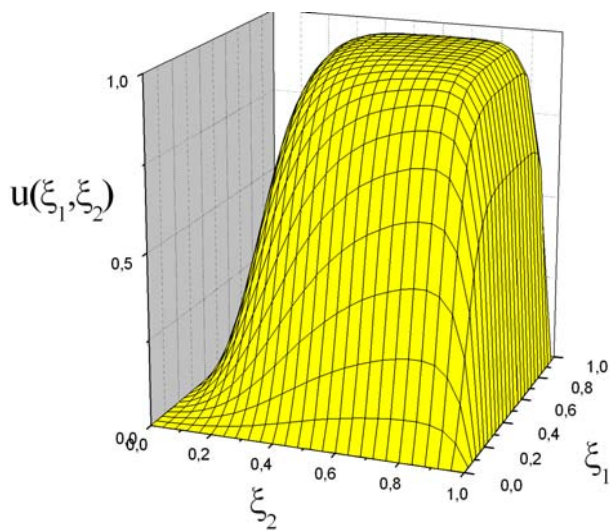

*Fig.2c* Steady-state distribution of the negative charged particles concentration ($v_1=3$ and $v_2=3$).